\documentclass[aps,prl,twocolumn,superscriptaddress,reprint,showpacs]{revtex4-2}
\usepackage{graphicx}
\usepackage{setspace}
\usepackage{amsmath}
\usepackage{amssymb}
\usepackage{color}
\usepackage{array}
\usepackage{subfigure}
\usepackage{hyperref}
\usepackage{float}
\usepackage{lipsum}

\usepackage[all]{xy}
\newcommand{\RN}[1]{%
  \textup{\uppercase\expandafter{\romannumeral#1}}%
}

\begin{document}
\title{High fidelity entanglement of neutral atoms via a Rydberg-mediated single-modulated-pulse controlled-PHASE gate}
\author{Zhuo Fu}
\affiliation{State Key Laboratory of Magnetic Resonance and Atomic and Molecular Physics, Innovation Academy for Precision Measurement Science and Technology, Chinese Academy of Sciences-Wuhan National Laboratory for Optoelectronics, Wuhan 430071, China}
\affiliation{School of Physical Sciences, University of Chinese Academy of Sciences, Beijing 100049, China}
\author{Peng Xu}
\affiliation{State Key Laboratory of Magnetic Resonance and Atomic and Molecular Physics, Innovation Academy for Precision Measurement Science and Technology, Chinese Academy of Sciences-Wuhan National Laboratory for Optoelectronics, Wuhan 430071, China}
\author{Yuan Sun}
\affiliation{Key Laboratory of Quantum Optics and Center of Cold Atom Physics, Shanghai Institute of Optics and Fine Mechanics, Chinese Academy of Sciences, Shanghai 201800, China}
\author{Yangyang Liu}
\affiliation{State Key Laboratory of Magnetic Resonance and Atomic and Molecular Physics, Innovation Academy for Precision Measurement Science and Technology, Chinese Academy of Sciences-Wuhan National Laboratory for Optoelectronics, Wuhan 430071, China}
\affiliation{School of Physical Sciences, University of Chinese Academy of Sciences, Beijing 100049, China}
\author{Xiaodong He}
\affiliation{State Key Laboratory of Magnetic Resonance and Atomic and Molecular Physics, Innovation Academy for Precision Measurement Science and Technology, Chinese Academy of Sciences-Wuhan National Laboratory for Optoelectronics, Wuhan 430071, China}
\author{Xiao Li}
\affiliation{State Key Laboratory of Magnetic Resonance and Atomic and Molecular Physics, Innovation Academy for Precision Measurement Science and Technology, Chinese Academy of Sciences-Wuhan National Laboratory for Optoelectronics, Wuhan 430071, China}
\author{Min Liu}
\affiliation{State Key Laboratory of Magnetic Resonance and Atomic and Molecular Physics, Innovation Academy for Precision Measurement Science and Technology, Chinese Academy of Sciences-Wuhan National Laboratory for Optoelectronics, Wuhan 430071, China}
\author{Runbing Li}
\affiliation{State Key Laboratory of Magnetic Resonance and Atomic and Molecular Physics, Innovation Academy for Precision Measurement Science and Technology, Chinese Academy of Sciences-Wuhan National Laboratory for Optoelectronics, Wuhan 430071, China}
\author{Jin Wang}
\affiliation{State Key Laboratory of Magnetic Resonance and Atomic and Molecular Physics, Innovation Academy for Precision Measurement Science and Technology, Chinese Academy of Sciences-Wuhan National Laboratory for Optoelectronics, Wuhan 430071, China}
\author{Liang Liu}
\email[email: ]{liang.liu@siom.ac.cn}
\affiliation{Key Laboratory of Quantum Optics and Center of Cold Atom Physics, Shanghai Institute of Optics and Fine Mechanics, Chinese Academy of Sciences, Shanghai 201800, China}
\author{Mingsheng Zhan}
\email[email: ]{mszhan@wipm.ac.cn}
\affiliation{State Key Laboratory of Magnetic Resonance and Atomic and Molecular Physics, Innovation Academy for Precision Measurement Science and Technology, Chinese Academy of Sciences-Wuhan National Laboratory for Optoelectronics, Wuhan 430071, China}

\date{\today}

\begin{abstract}
Neutral atom platform has become an attractive choice to study the science of quantum information and quantum simulation, where intense efforts have been devoted to the entangling processes between individual atoms.
For the development of this area, two-qubit controlled-PHASE gate via Rydberg blockade is one of the most essential elements.
Recent theoretical studies have suggested the advantages of introducing non-trivial waveform modulation into the gate protocol, which is anticipated to improve its performance towards the next stage.
We report our recent experimental results in realizing a two-qubit controlled-PHASE($C_Z$) gate via off-resonant modulated driving(ORMD) embedded in two-photon transition for Rb atoms.
It relies upon a single modulated driving pulse with a carefully calculated smooth waveform to gain the appropriate phase accumulations required by the two-qubit gate. Combining this $C_Z$ gate with global microwave pulses, two-atom entanglement is generated with the raw fidelity of 0.945(6).
Accounting for state preparation and measurement (SPAM) errors, we extract the entanglement operation fidelity to be 0.980(7).
Our work features completing the $C_Z$ gate operation within a single pulse to avoid shelved Rydberg population, thus demonstrate another promising route for realizing high-fidelity two-qubit gate for neutral atom platform.
\end{abstract}
\maketitle

Neutral atoms have long been deemed as an essential platform in the study of quantum information \cite{Saffman2010, Saffman2016, Henriet2020, Morgado2021} and quantum simulation \cite{Browaeys2020, Scholl2021, Ebadi2021}, and recent rapid progress has revealed that cold atoms in optical traps serve as an ideal choice of qubits, where Rydberg blockade \cite{Jaksh2000, Isenhower2010, Wilk2010} serves as the backbone for the entangling processes between individual atoms.
So far, intense efforts have been devoted to a wide range of experimental topics in this area, including the increment of number and type of qubits in array format \cite{Kumar2018, Mello2019}, the enhancement of entangled state size \cite{Omran570}, and the improvement of quantum gate performance \cite{Ebert2015, Lukin2019, Graham2019}. These developments clearly demonstrate the promising potential of neutral atom qubit and pave the way for future applications.

Amid many pressing tasks, an imminent challenge is to enhance the two-qubit gate fidelity towards the requirement of Noisy Intermediate-Scale Quantum technology (NISQ) under realistic experimental conditions \cite{Weiss2017, Henriet2020, Morgado2021}.
Encouragingly, the fidelity of entangling two strontium atoms directly from ground state using Rydberg blockade has reached high fidelity, which utilize the long-coherence Rabi oscillations of the ground-to-Rydberg-state transitions \cite{Madjorov2020}.
However, the Rydberg-mediated two-qubit quantum gate demonstrated in current experiments, either $C_Z$ gate or controlled-NOT gate, still yield less than satisfying fidelities, even with suppressed phase noise of the Rydberg-excitation lasers, long-coherence Rabi oscillations and optimized experimental conditions \cite{Ebert2015, Jau2015, Zeng2017, Picken2018, Graham2019}.
One major reason for the infidelity is that atoms of superposition states are excited to the Rydberg state and are held on for a certain time, thus suffer from the severe decoherence between the ground state and the Rydberg state which is indicated by the ground-to-Rydberg-state Ramsey oscillations \cite{Liu2021}. To realize higher fidelity two-qubit quantum gate, it is necessary to reduce the Ramsey decoherence induced infidelity, such as increasing the ratio of Ramsey coherence time to gate time as done by Ref. \cite{Lukin2019}.

Alternatively, it is possible to develop new schemes of $C_Z$ gate to avoid shelved Rydberg populations. Inspired by the controlled-PHASE gate protocol for single-photon ground-Rydberg transition with non-trivial smooth waveform modulation methods \cite{Sun2020} which provides the feasibility of driving both qubit atoms with the same control field, here, we develop a new method of $C_Z$ gate for two-photon transition via Single-modulated-pulse Off-Resonant Modulated Driving (SORMD).
Using global Rydberg excitation of two $^{87}$Rb atoms within the Rydberg blockade region, $C_Z$ gate is realized via SORMD, where atoms are continuously drived in the ground-Rydberg transitions. We benchmark the gate fidelity by measuring the raw fidelity of two-atom entanglement to be $\mathcal{F}$=0.945(6).
Correcting for state preparation and measurement (SPAM) errors, we extract the fidelity of entanglement operation to be $\mathcal{F}^c$ = 0.980(7).

\begin{figure}[t]
% \centering
\includegraphics[width=7cm]{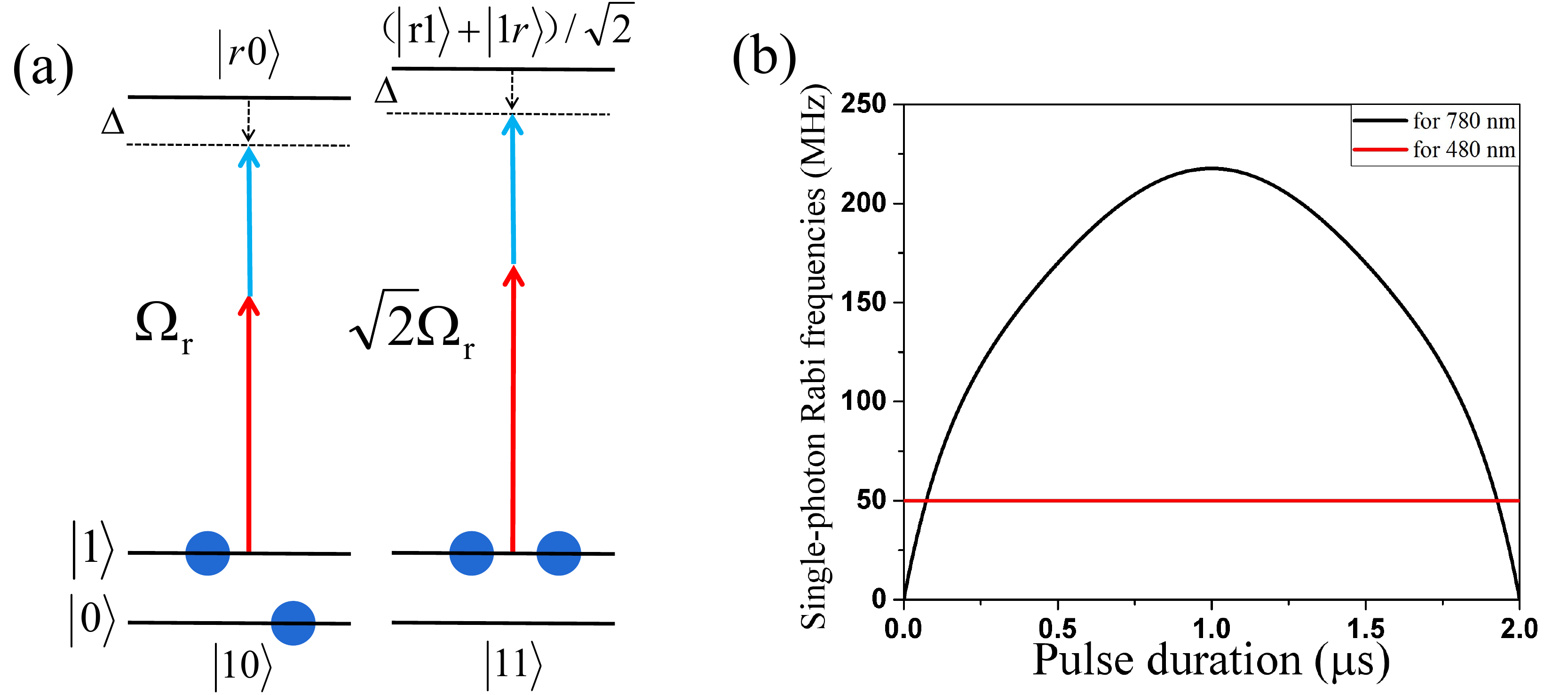}
\includegraphics[width=7cm]{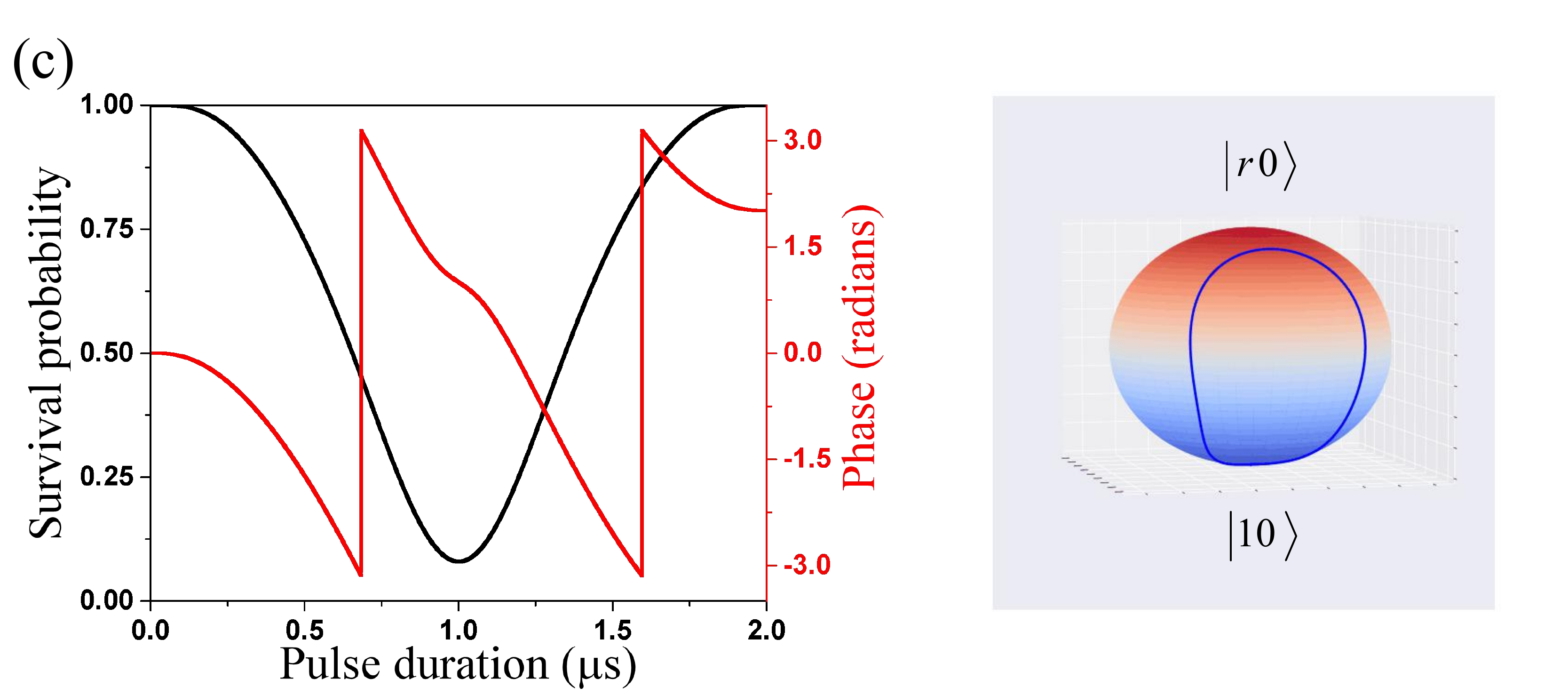}
\includegraphics[width=7cm]{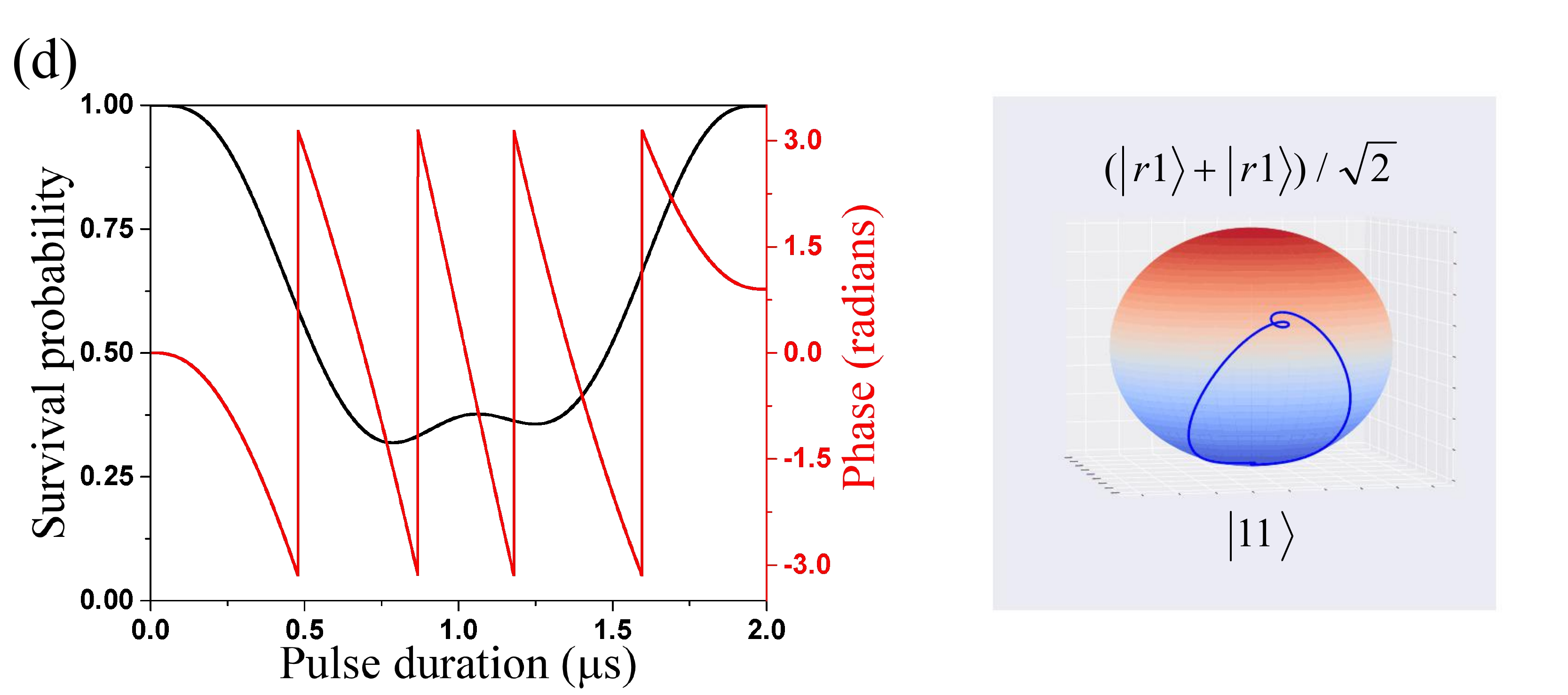}
\caption{Protocol of $C_Z$ gate via Single-modulated-pulse Off-Resonant Modulated Driving(SORMD).(a)Level diagram for two-photon Rydberg excitation. With Rydberg blockade, two atoms in $|11\rangle$ state will be excited to $(1/\sqrt{2})(|1r\rangle+|r1\rangle)$ with an enhanced effective Rabi frequency. (b) Calculated waveform of SORMD pulse. SORMD only modulates the amplitude of 780-nm laser with a waveform of a linear combination of Bernstein basis polynomials, and keeps 480-nm laser at a constant value.  (c) The population and phase dynamics of $|01\rangle$ and $|10\rangle$ states. The population returns back with accumulated phase $\phi_{01} = \phi_{10} = 2.012 $. (d) The population and phase dynamics of $|11\rangle$. The population also returns back but accumulates phase $\phi_{11}= 0.8997$. The Rydberg excitation laser is far off resonant for $|00\rangle$ state, thus $\phi_{00}$ can be ignored. After SORMD pulses, $\phi_{00} - \phi_{01} - \phi_{10} + \phi_{11} = -0.995 \pi$. }
\label{fig:basic_0}
\end{figure}

In our experiment, SORMD method resorts to specially tailored waveform of 780-nm Rydberg-excitation laser to gain appropriate phase accumulations under the presence of Rydberg blockade, as illustrated in Fig. \ref{fig:basic_0}. Specifically, SORMD generates phases of the two-qubit computational basis states $|00\rangle$,  $|01\rangle$,  $|10\rangle$ and $|11\rangle$ as
\begin{equation}
\label{eq:Cz}
\phi_{00} - \phi_{01} - \phi_{10} + \phi_{11} = \pm \pi,
\end{equation}
The two-photon Rydberg-excitation configuration provides extra controlling laser and ac Stark shifts which significantly differs from single-photon transition \cite{Sun2020}. For experimental simplicity, we set 480-nm Rydberg excitation laser as the constant driving laser while analysis and calculate an amplitude modulated waveform for 780-nm laser that starts and ends at zero. Without loss of generality, we choose a representation of the waveform in terms of a linear combination of basis polynomials:
\begin{equation}
\Omega_r (t)/2\pi = \sum_{\nu=1}^{4} \beta_\nu \big(b_{\nu, n}(t/T_g) + b_{n-\nu, n}(t/T_g) \big),
\end{equation}
where $b_{\nu, n}$ is the $\nu$th Bernstein basis polynomials of degree $n$. The advantages of such a representation include its smoothness, compatibility with numerical optimization procedure and without a long tail. Under ideal conditions, the ultimate limit for the gate fidelity is the spontaneous lifetime of Rydberg levels and the residual thermal motion of qubit atoms. Here, for our experiment we choose the following set of parameters after optimization of the waveform:$\beta_1=206.4 \text{MHz}, \beta_2=90.1 \text{MHz}, \beta_3=300.5 \text{MHz}, \beta_4=195.97 \text{MHz}$, intermediate detuning of $\Delta$/2$\pi$ = -5687 MHz  with $T_g = 2$ $\mu$s, Rabi frequency of 480-nm laser at 50 MHz and the an overall net two-photon detuning at $\delta$/2$\pi$ = 1.50 MHz. Theoretically, the fidelity of two-atom entanglement is up to 0.99 with this method.

\begin{figure}[t]
\includegraphics[width=8.5cm]{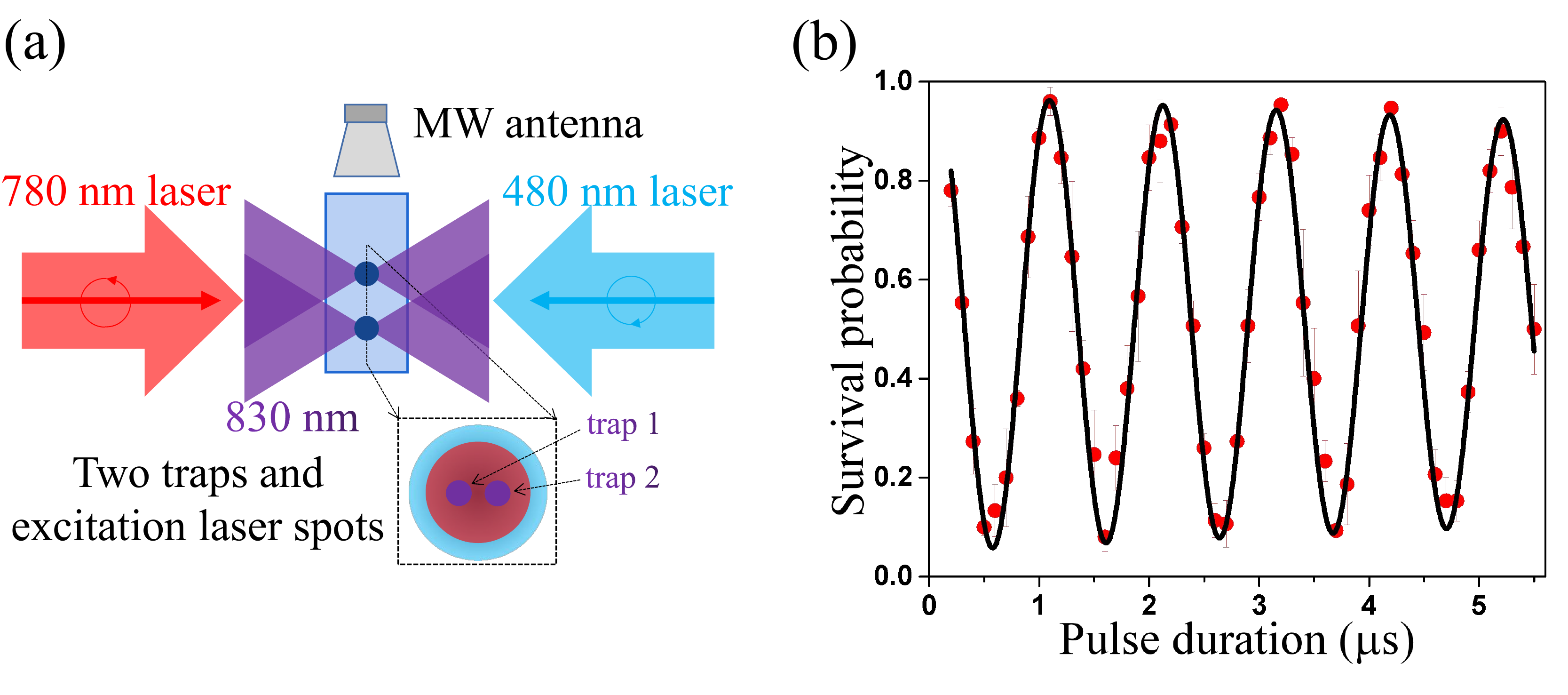}
\includegraphics[width=8.5cm]{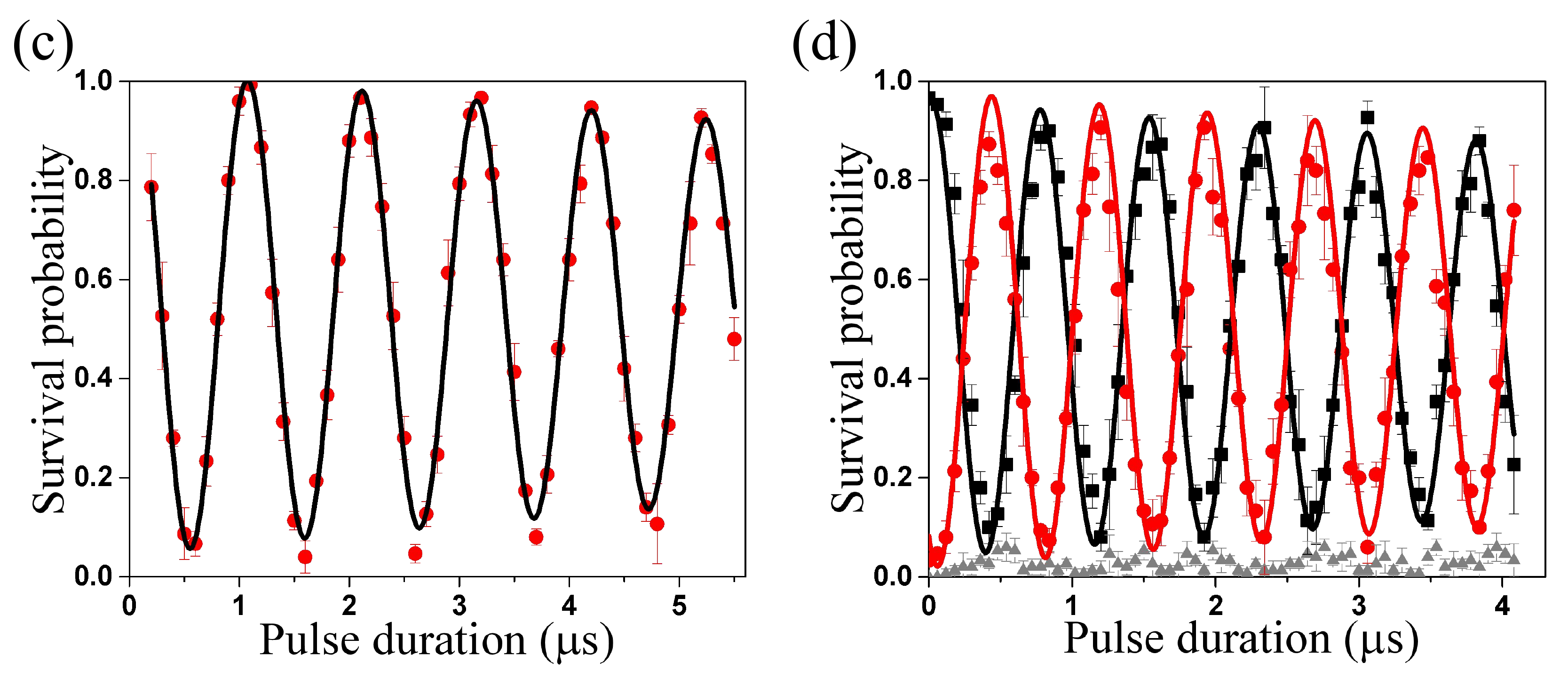}
\caption{Experimental setup and resonant two-photon ground-to-Rydberg-state Rabi oscillations. (a) Relevant lasers' configuration. Single atom Rabi oscillations between  $|1\rangle$ and $|r\rangle$ in trap-1 (b) and in trap-2 (c). Red dots are measured survival probability of atoms after Rydberg excitation which repeat for 150 times. Black solid curves are damped sinusoidal fits, which gives effective Rabi frequency of  2$\pi$$\times$0.96(1) MHz and a deacy time of 23(2) $\mu$s. (d) Two-atom collective Rabi oscillations between $|11\rangle$ and $(1/\sqrt{2})(|1r\rangle+|r1\rangle)$. Black squares are measured probability of atoms survived in both traps, red circles are measured probability of atoms only survived in one trap, while grey triangles are measured probability of atoms lost in both traps. The fitted Rabi frequency is 2$\pi$$\times$1.35(2) MHz and a decay time about 20 $\mu$s.}
\label{fig:basic_1}
\end{figure}

The main experimental setup is shown in Fig. \ref{fig:basic_1}. In our experiment, two $^{87}$Rb atoms are trapped in two optical tweezers, which are generated by the tightly focused 830 nm laser, with the beam waist at focal plane 1.2(1) $\mu$m and separated by 3.6(1) $\mu$m. The temperature of single-atoms is about 5.2 $\mu$k in a 50 $\mu$k trap after we applying polarization-gradient cooling and adiabatic cooling. The qubits are encoded into the hyperfine ground state of $^{87}$Rb atom,with $\vert$1$\rangle$=$\vert$5$S_{1/2}$,F=2,$m_F$=0$\rangle$ and $\vert$0$\rangle$=5$S_{1/2}$,F=1,$m_F$=0$\rangle$.
We initialize all qubits by preparing atoms in $\vert$1$\rangle$ through an optical pumping procedure, with an efficiency of 99.2(2)$\%$.
The single-qubit operation is realized by 6.8 GHz microwave radiation and the Rabi oscillation frequency is about 2$\pi$$\times$33 kHz. We detect the atomic state by applying a resonant laser pulse which blows away of atoms in state $\vert$1$\rangle$. The detection efficiency for state  $\vert$1$\rangle$ is about 0.7$\%$ per atom, and atom preservation probability is about 98.9$\%$ in one trap.

\begin{figure}[t]
\includegraphics[width=8.5cm]{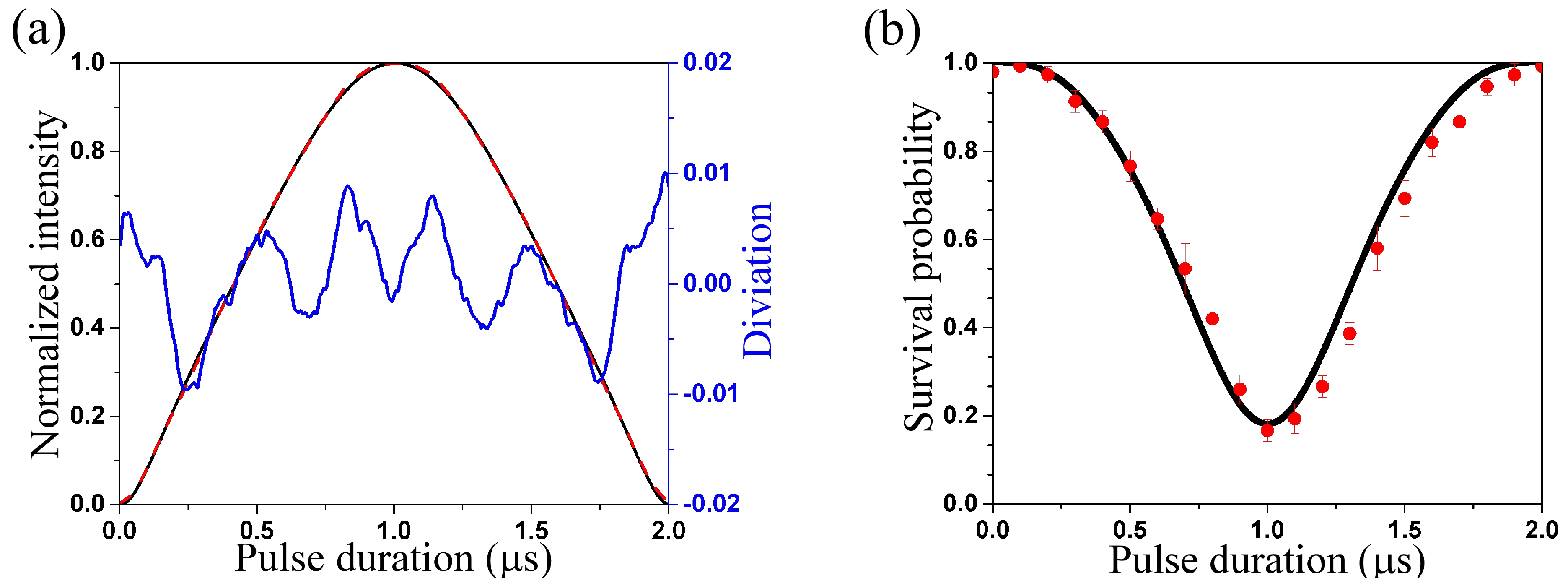}
\includegraphics[width=8.5cm]{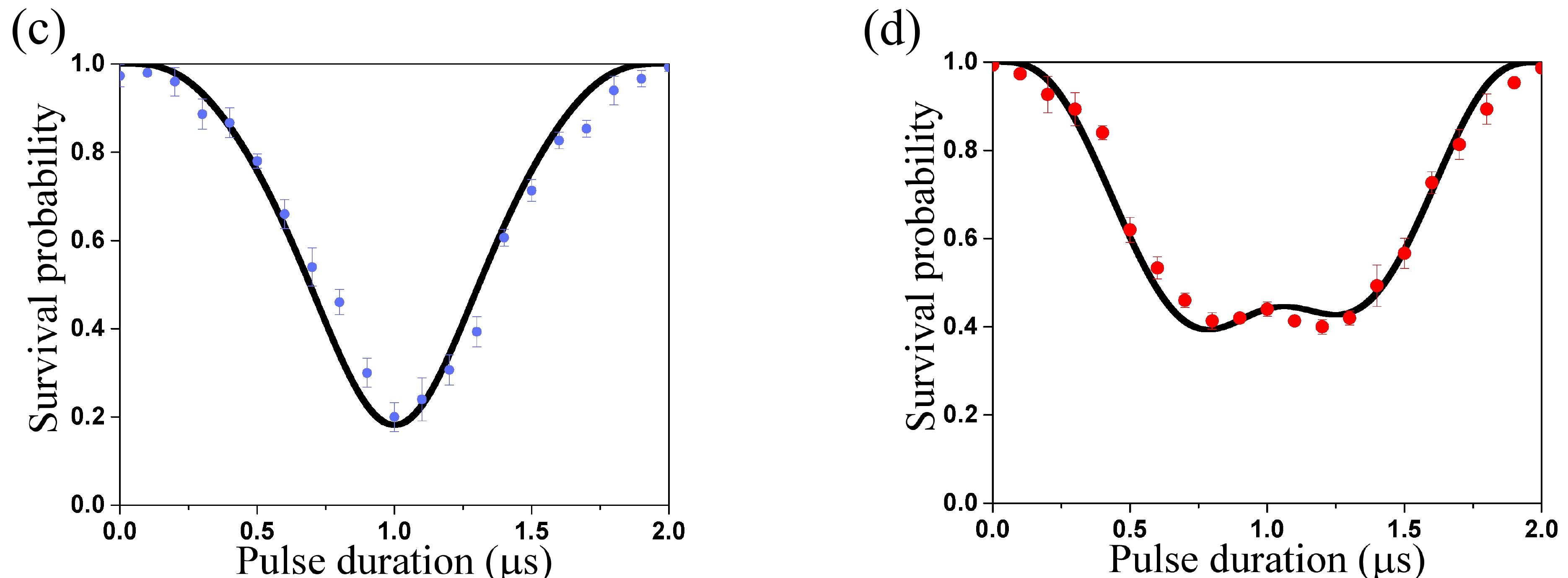}
\caption{Waveform of 780-nm laser pulse and population evolutions of SORMD. (a) Calculated (black solid line) and generated (red dashed line) intensity waveform of 780 nm pulse for SORMD. Blue line is the experimental deviation from theoretical value. Population evolution of single atom with $|1\rangle$ state in trap-1 (b) and in trap-2 (c) using SORMD method. (d) Population evolution of $|11\rangle$ state using SORMD method. The maximal single-photon Rabi frequency of 780-nm laser and 480 nm laser are adjusted to 217.6 MHz and 50 MHz, respectively. The circles are experimental results with 150 repeations. Solid curves are theoretical predictions considering the 88.9\% detection efficiency of Rydberg state. }
\label{fig:basic_2}
\end{figure}

We excite single-atoms from $|1\rangle$ state to the Rydberg state $|r\rangle$=$\vert$79$D_{5/2}$,$m_j$=5/2$\rangle$ using a two-photon transition with counter-propagating 780-nm ($\sigma^+$) laser and 480-nm ($\sigma^+$) laser.  Several technical improvements have been done to achieve long-coherence Rabi oscillations of the ground-to-Rydberg-state transitions, such as suppress the Rydberg-excitation lasers' phase noise around 750 kHz for about 30 dB with a high-finesse cavity, reduce Rydberg-excitation lasers' linewidth and frequency fluctuations, increase of single-photon Rabi frequnecies of $\Omega_{480}$, shield stray electric field and so on \cite{Zeng2018, CPB2021, Liu2021}. In our experiment, we adjust the single-photon Rabi frequencies to be $\Omega_{780}\simeq 2\pi \times$ 217 MHz and  $\Omega_{480}$$\simeq$2$\pi$$\times$50 MHz) with 780-nm laser detuned from 5$P_{3/2}$ of $\Delta$=2$\pi$$\times$-5687 MHz.
From the measured Rabi oscillations between $|1\rangle$ and $|r\rangle$ in Fig.\ref{fig:basic_2}, we deduce a 1/e decay time of $\tau$=23(2) $\mu$s, Rydberg-excitation efficiency $P_{RE}$=98.9(3)$\%$ and Rydberg-detection efficiency $P_{RD}$=88.9(9)$\%$. Using a numerical simulation with qutip \cite{qutip}, we conclude that the decay of Rabi oscillation is mainly caused by the atom experienced laser intensity fluctuation of 780-nm laser. Next, we come to collectively excite two atoms by enlarging the beam waist of 780-nm laser to 7.8(3) $\mu$m and 480-nm laser to 8.3(5) $\mu$m to cover the two traps as uniform as possible. Two optical traps are horizontally and symmetrically placed in the center of the global excitation laser beams as Fig. 2(a) shown. Since the laser field globally couples two atoms from $|11\rangle$ to $|W\rangle =(1/\sqrt{2})(|1r\rangle+|r1\rangle$), we observe an enhanced Rabi frequency of $\sqrt{2}\Omega_1\simeq 2\pi \times$1.35 MHz [see Fig. \ref{fig:basic_1}(d)] with Rydberg blockade. The residue probability of atoms lost in both traps can be attributed to atom loss in single traps and imperfect Rydberg blockade.

To successfully implement SORMD, the most important thing is to coherently and precisely drive the atoms to evolve as calculated. Thus, we first adjust single-photon Rabi frequencies of Rydberg-excitation lasers to the expected values with deviations less than 1\%.
For 780-nm laser, different laser powers will cause different AC-stark shifts of the Rydberg excitation peak. By linear fitting Rydberg peaks with laser powers, we can determine 780-nm laser of 59.2 $\mu$W correspond to $\Omega_{780} = 2\pi \times 217(2)$ MHz.
Besides, we measure the Ramsey fringes of state $|0\rangle$ to $|1\rangle$ with different durations of 780 nm pluses in-between the $\frac{\pi}{2}$ pulses, the calculated $\Omega_{780}$ from the shifted phases of Ramsey fringes agrees well with previous result. $\Omega_{480}$ is then determined to be $2\pi \times 50.0(5)$ MHz of 120 mW laser power by measuring the Rabi oscillation frequency between $|1\rangle$ and $|r\rangle$. Second, we achieve the differences of $\Omega_{780}$ ($\Omega_{480}$) between two traps to be less than 0.5\% (0.7\%). This is done by fine tuning the position of 780-nm laser and 480-nm laser with piezoelectric(Thorlabs, POLARIS-K1S2P) drived mirrors. Third, to generating SORMD waveform of 780-nm laser, we modulate the amplitude of the RF signal which is used to drive the acousto-optic modulator (AOM) via the input port of a mixer(Mini-circuits ZFM-3-S+). The generated waveforms is in good agreement with the theoretical calculation and the deviation is less than 1\%, as shown in Fig.\ref{fig:basic_2}(a). When we finished these preparations, we excite the atoms with SORMD and measure the population of atoms in state $\vert$1$\rangle$ as FIG. \ref{fig:basic_2} shows. Compared with the theoretical calculations, the evolution of the population curves is within expectations and the imperfections may be caused by the excitation frequency and single-photon Rabi frequencies fluctuations in two traps.

\begin{figure}[t]
\includegraphics[width=8.0cm]{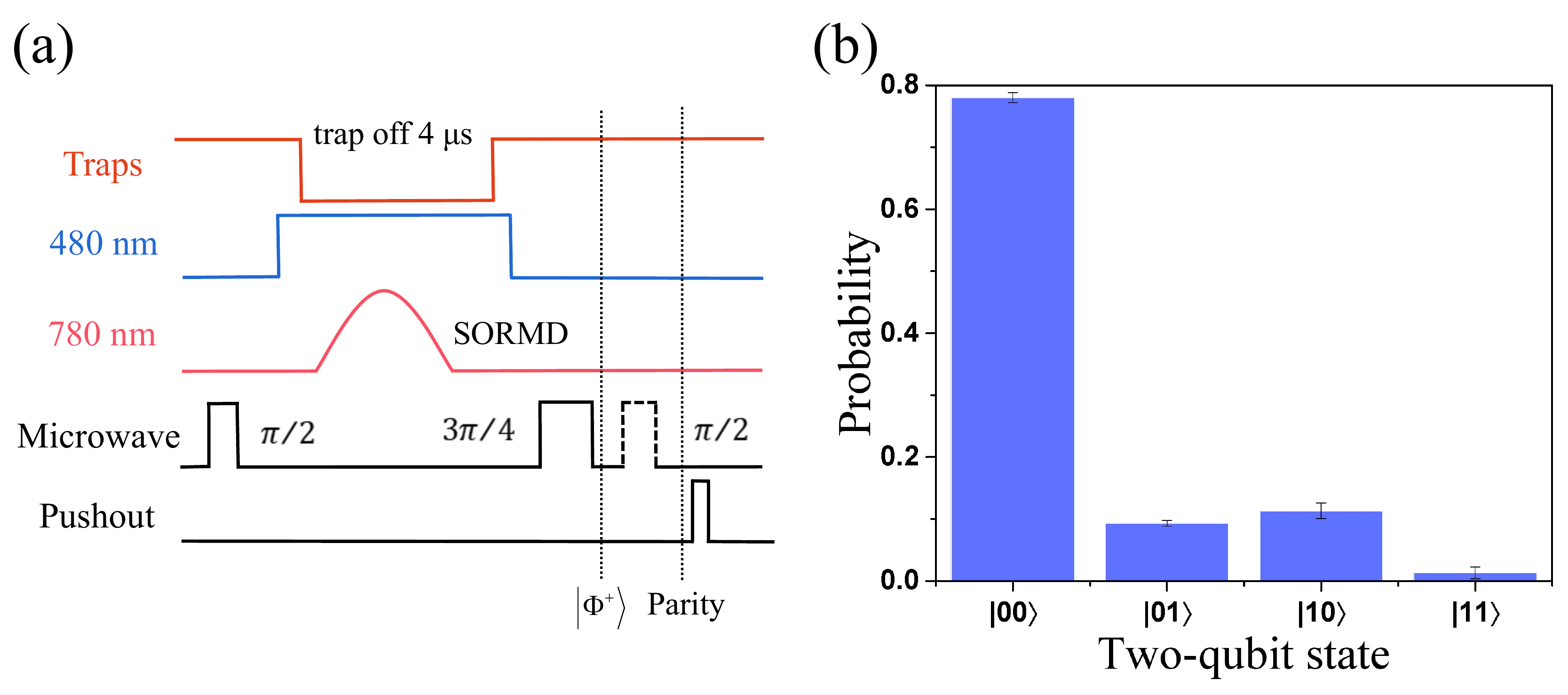}
\includegraphics[width=8.5cm]{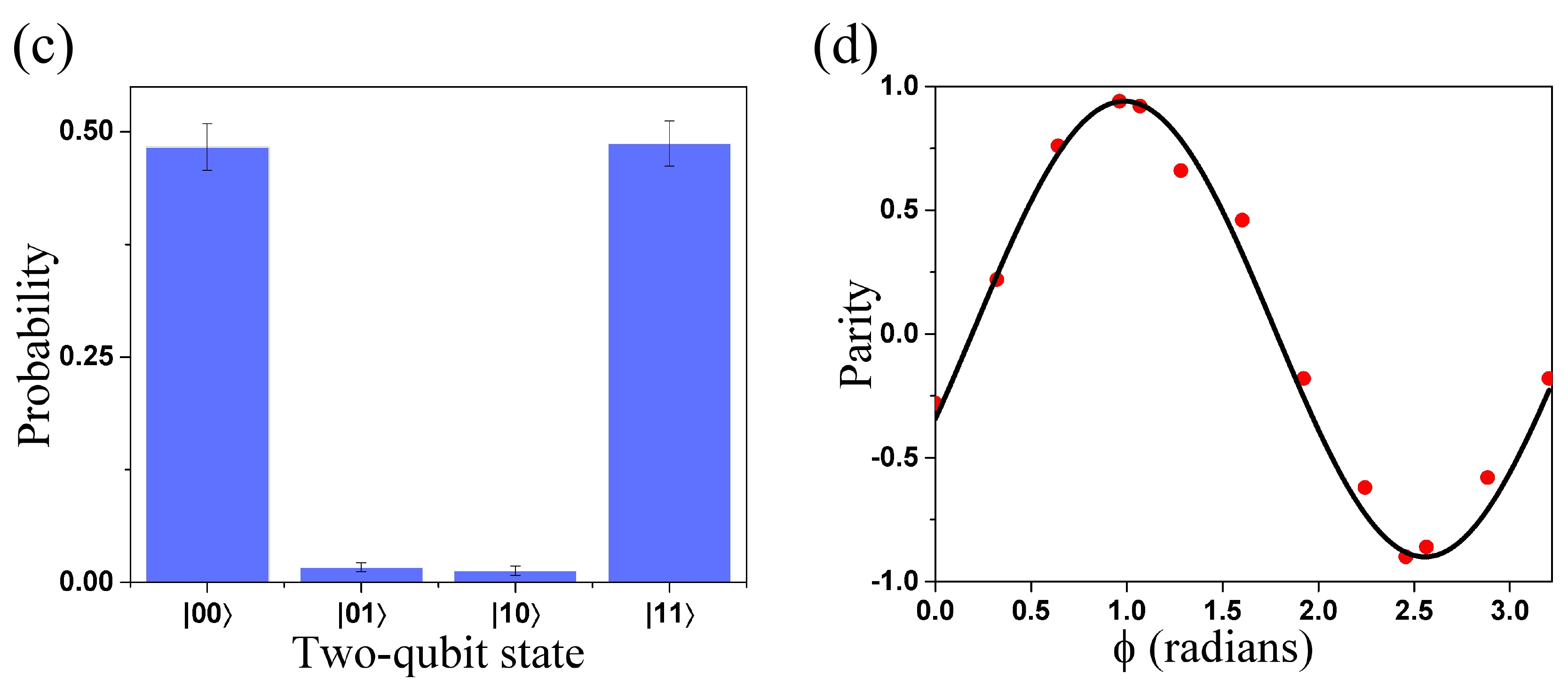}
\caption{$C_Z$ gate realized by SORMD method. (a) The main pulse sequence for preparing Bell states $\vert\Phi^+\rangle$. All the operations are global. (b)  Measured population of two-qubit state without SORMD. (c) Bell state population. Raw measurements gives ($P_{00}$+$P_{11}$)/2=0.485(4). (d) The measured amplitude of parity oscillation is 0.92(1), and the raw fidelity of Bell state is $F_{Bell}$=0.485+0.46=0.945(6).}
\label{fig:basic_3}
\end{figure}

We illustrate that SORMD can realize the $C_Z$ gate by using it to create two atoms Bell states $|\Phi^+\rangle=(1/\sqrt{2})(|00\rangle+|11\rangle$), and the pulse sequence is showed in FIG. \ref{fig:basic_3}. We first initialize the two atoms in state $|11\rangle$ and then apply a global microwave $\frac{\pi}{2}$ pulse. After we apply a 2 $\mu$s SORMD, we use a global microwave $\frac{3\pi}{4}$ pulse to produce the Bell state $\vert\Phi^+\rangle$. In addition, we scan the phase of the last global microwave $\frac{3\pi}{4}$  pulse to compensate the relative phase of the $C_Z$  gate. As a comparison, we repeat the same pulse sequence but without SORMD, the population difference clearly shows that SORMD creats a $C_Z$ gate.[FIG. 4(b) and 4(c)]. To benchmark the gate fidelity, we measure the coherence of Bell states by applying a global microwave  $\frac{\pi}{2}$ pulse with a variable phase. Fig. \ref{fig:basic_3} shows the resulting data of Bell state population and the parity oscillation, and it gives ($P_{00}$+$P_{11}$)/2=0.485(4), partiy amplitude C=0.460(5) and $F_{Bell}$=0.485+0.46=0.945(6). Relying on the method of state measurement in our experiment, the atom will be classified as state $\vert$1$\rangle$ when it is lost due to background collision or finite atomic temperature. To correct the atom loss caused the overestimation of state $\vert$11$\rangle$ in Bell state preparation, we measure the population in which we disable the pushout beam. The corrected data gives ($P_{00}$+$P_{11}$)/2$\ge$0.475(4). Both Bell state measurement and the parity oscillation measurement should correct for state preparation and measurement(SPAM) errors which is about 2.6$\%$ per atom. Referring to the methods about correction for SPAM errors in \cite{Lukin2019}, Bell state population corrected as ($P_{00}^{SPAM}$+$P_{11}^{SPAM}$)/2$\ge$0.495(5), the parity oscillation corrected as $C^{SPAM}$=0.485(6) and the Bell state fidelity corrected as $F_{Bell}^{SPAM}$=0.980(7).

\begin{center}
\textbf{Table 1}~~Calibrating the fidelity of  (1/$\sqrt{2}$)($\vert$00$\rangle$+$\vert$11$\rangle$) generated by SORMD.\\
\setlength{}{
\begin{tabular}{cccc}
\hline \hline
& Raw data \; & Lower bound \; & Corrected  \\
\hline
Population & 0.970(8) & 0.950(8) & 0.990(10) \\
Coherence & 0.920(10) & 0.920(10) & 0.970(12) \\
\hline
Fidelity & 0.945(6) & 0.935(6) & 0.980(7) \\
\hline \hline
\end{tabular}}
\end{center}

%\section{Conclusion and Outlook}
In conclusion, we have experimentally implemented a new category of two-qubit controlled-PHASE gate based upon Rydberg blockade effects. In our method, atoms are continuously drived in the ground-Rydberg transitions to make full use of long-coherence Rabi oscillations instead of the shorter-coherence Ramsey oscillations. We are also looking forward to a few other future refinements, including the search for a faster gate operation, further suppression of population leakage, stronger robustness against environmental noises, and more user-friendly parameter setting. Error correction mechanism \cite{PhysRevLett.117.130503} for our gate protocol is also part of the long term goal. Our aim is to help the translation of high-quality ground-Rydberg coherence into high-fidelity controlled-PHASE gate via Rydberg blockade effect.

\begin{acknowledgements}
Zhou Fu, Peng Xu and Yuan Sun contribute equally. 
The authors gratefully acknowledge the funding support from the National Key R\&D Program of China (under contract Grant No. 2016YFA0302800 and No. 2016YFA0301504), the Youth Innovation Promotion Association CAS No. 2017378, the National Natural Science Foundation of China under Grant No. 11774389, and the Strategic Priority Research Program of the Chinese Academy of Sciences under Grant No. XDB21010100. The authors gratefully thank Zongyuan Xiong for support of electrical circuits, Qunfeng Chen for technique support of high finess cavity.
\end{acknowledgements}

\bibliographystyle{apsrev4-2}

%\renewcommand{\baselinestretch}{1}
%\normalsize

%\clearpage%
%\phantomsection%
%\addcontentsline{toc}{chapter}{\numberline{}{Bibliography}}%
\bibliography{ORMD_ref}

\clearpage

\end{document}